\documentclass[prl,twocolumn,superscriptaddress,letterpaper, showpacs,amsmath,amssymb,floatfix,nofootinbib]{revtex4-1}
\usepackage[usenames,dvipsnames]{color}
\usepackage{graphicx,epstopdf}
\usepackage[bookmarks]{hyperref}
\usepackage[normalem]{ulem}

\begin{document}
%%%%%%%%%%%%%%%%%%%%%%%%%%%%%%%%%%%%%%%%%%
%\title{Microscopic study of K-Selection Rule Violation in $^{182}$W K=10$^+$ Isomer Decay}

%\author{C. R. Praharaj}
%\email{crp@iopb.res.in}
%\affiliation{Institute of Physics, Bhubaneswar - 751005, India}

%\author{S. Bhoi}
%\email{swagatikabhoi66@gmail.com}
%\affiliation{School of Physics, Sambalpur University, Jyoti Vihar, Burla-768019, India}

%\author{Z. Naik}

%\affiliation{School of Physics, Sambalpur University, Jyoti Vihar, Burla-768019, India}

%\author{S. K. Ghorui}

%\affiliation{School of Physics and Astronomy, S.J.T University, Shanghai 200240, China}

%\author{S. K. Patra}

%\affiliation{Institute of Physics, Bhubaneswar - 751005, India}

%\date{\today}

%%%%%%%%%%%%%%%%%%%%%%%%%%%%%%%%%%%%%%%%%%%%%%%%%%%5
\title{Microscopic study of K-Selection Rule Violation in $^{182}$W K=10$^+$ Isomer Decay}

\author{C. R. Praharaj*}
%\email{crp@iopb.res.in}
\affiliation{Institute of Physics, Bhubaneswar - 751005, India}

\author{Swagatika Bhoi*}%$^{1}$}
\affiliation{School of Physics, Sambalpur University, Jyoti Vihar, Burla-768019, India}

\author{Z. Naik}

\affiliation{School of Physics, Sambalpur University, Jyoti Vihar, Burla-768019, India}

\author{S. K. Ghorui}

\affiliation{School of Physics and Astronomy, S.J.T University, Shanghai 200240, China}

\author{S. K. Patra}

\affiliation{Institute of Physics, Bhubaneswar - 751005, India}
%\affiliation{\it $^{3}$Department of Physics, University of Maryland, College Park, Maryland 20742, USA}
\date{\today}
%%%%%%%%%%%%%%%%%%%%%%%%%%%%%%%%%%%%%%%%%%%
\begin{abstract}
In this paper we study the microscopic mechanism for the retarded decay of K-isomers to lower K bands. We do angular momentum projection from suitable intrinsic states. The retardation arises from poor overlap between the low K and high K bands in the integral over Euler angles. Deformed HF and angular momentum projection calculations are done for the decay of the $K=10^+$ isomer band to the ground band of $^{182}W$. K-mixing is unimportant and the K quantum numbers of the bands are quite good. There is significant difference in the reduced matrix elements of transition operators in our formalism and that in the rotational model. Angular momentum projection gives J-selection rule but there is no K selection rule for reduced matrix elements of electromagnetic multipole operators.  Thus, E2 and M1 transitions from the K isomer to the ground band of $^{182}W$ are finite but retarded in angular momentum projection theory, as in experiments. This provides a theoretical basis for the study of K-isomers and their decay modes. Quantitative results are presented. The microscopic model gives J-selection rule and angular momentum conservation for combined matter and radiation systems.

\end{abstract}

\maketitle
\footnotetext [1]{crp@iopb.res.in}
\footnotetext [2]{swagatikabhoi66@gmail.com}
% % \linenumbers

%%%%%%%%%%%%%%%%%%%%%%%%%%
\paragraph{\label{sec:intro}\bf{Introduction:}} 
%%%%%%%%%%%%%%%%%%%%%%%%%%
The problem of the decay of K isomers and the associated K selection rule and its violation in deformed nuclei have been subjects of much discussion. Many cases of K violating gamma-ray transitions are known experimentally \cite{kondev, regan, chowdhury, cullen, hayes, walker}. No consistent explanation is available for the finite but retarded transitions from K isomers to lower K bands seen experimentally. The rotational model forbids such transitions (the K-selection rule) \cite{bohr,alder} and in many discussions various ad hoc assumptions are made to explain the violation of K selection rule \cite{cullen}. But a proper quantitative understanding of the mechanism of the K-selection rule violation in the framework of nucleon-nucleon residual interaction is still lacking. One needs a refined model at the microscopic level to understand properly the low multipole decay modes of K isomers.

\begin{figure*}[ht]
\begin{center}
\includegraphics[width=1.3\columnwidth]{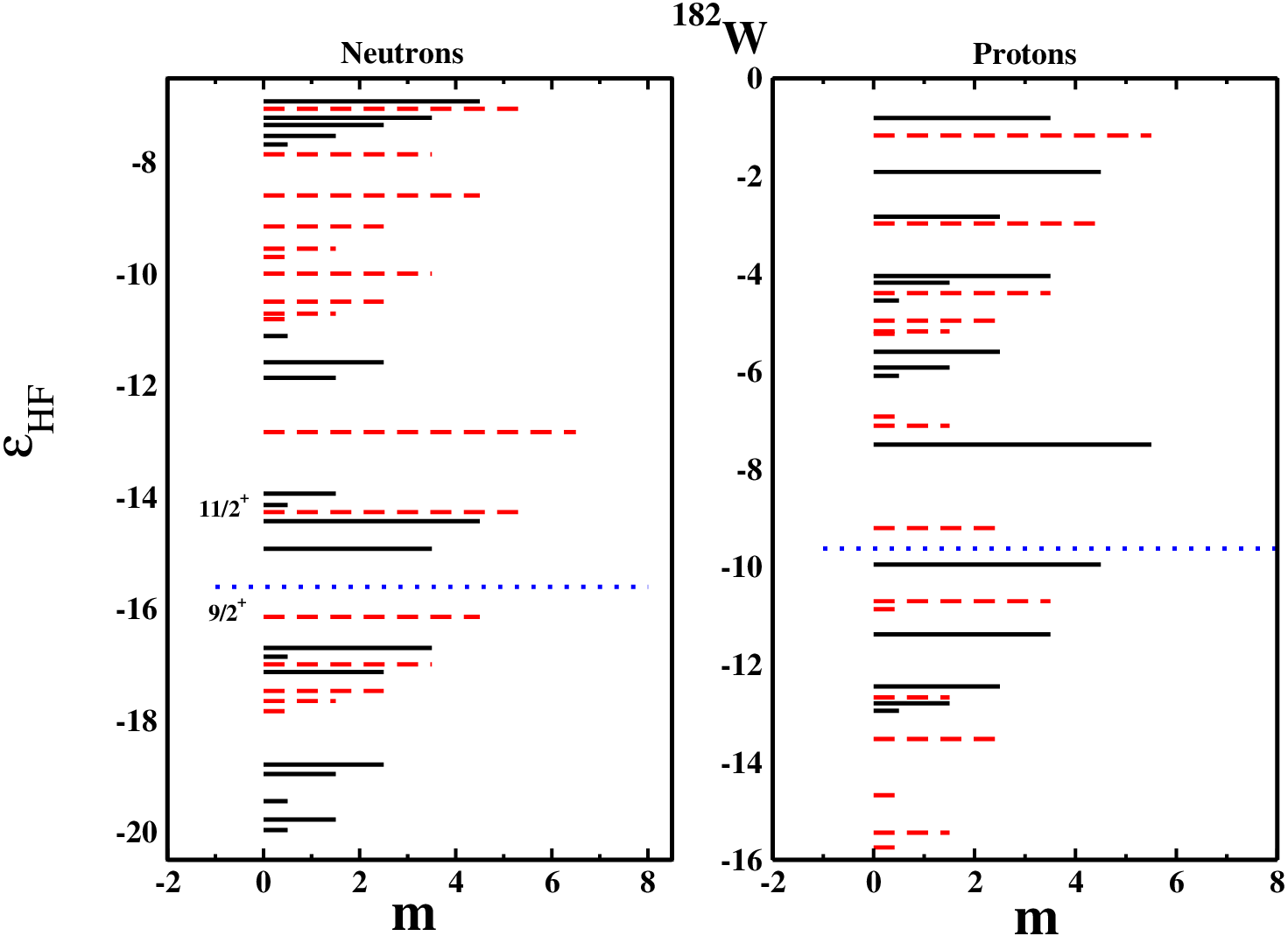}
\end{center}
\caption{Energy of HF orbits of $^{182}$W in MeV. The m value of an orbit is indicated by its length. Each orbit is doubly degenerate. The Fermi levels of neutrons and protons are indicated (dotted). The +ve parity orbits are in dashed red and the -ve parity orbits are in solid black.}
\label{fig1}
\end{figure*}
Peierls and Yoccoz \cite{peierls} showed more then 60 years ago that one needs angular momentum projection from deformed intrinsic states by applying rotation operator and integrating over the Euler angles in order to obtain physical states of good angular momentum J \cite{peierls, ripka, praharaj} (An intrinsic state of a deformed nucleus being a wave packet and a superposition of various J states, angular momentum projection is essential). We show in this work that a consistent application of the Peierls-Yoccoz procedure gives J selection rule for reduced matrix elements for electromagnetic transitions among states of various bands, but there is no K selection rule forbidding J allowed transitions between bands. In the following we briefly describe the deformed HF and angular momentum projection procedures and present results for deformations, energy spectra and electromagnetic transitions for the ground and the $\mu$-second K=10$^+$ band of $^{182}W$. It is found that E2 and M1 transitions from K=10$^+$ bandhead to the ground band of $^{182}$W are finite but retarded, in violation of K-selection rule, and in agreement with experimental trends \cite{kondev, regan}. We thus obtain a theoretical explanation of K-isomer decay at the microscopic level, without phenomenological assumptions or K mixing.

\paragraph{\bf{HF Theory and Angular Momentum Projection:}}
We use two major shells for protons and neutrons each with a surface delta residual interaction \cite{faessler} in our study. The model space consists of 1g$_{7/2}$, 2d$_{5/2}$, 2d$_{3/2}$, 3s$_{1/2}$, 1h$_{11/2}$, 1h$_{9/2}$, 2f$_{7/2}$, 1i$_{13/2}$, and 2g$_{9/2}$ with single-particle energies -6.92, -5.30, -3.58, -3.298, -4.376, 1.0, 2.0, 3.0 and 5.5 MeV respectively for protons. For neutrons we have 1h$_{9/2}$, 2f$_{7/2}$, 2f$_{5/2}$, 3p$_{3/2}$, 3p$_{1/2}$, 1i$_{13/2}$, 1i$_{11/2}$, 2g$_{9/2}$, and 1j$_{15/2}$ with single particle energies -10.943, -11.629, -8.407, -8.739, -7.776, -9.494, -4.049, -3.485 and -0.95. The force strength of the residual interaction (surface delta) is taken as 0.15 MeV which is appropriate for this calculation in extended model space. The band structure for a broad range of rare-earth nuclei are well described with this model space and interaction. $^{182}$W has neutron and proton numbers above the midshell and we use a fairly large model space to describe its spectra. The HF procedure is based on variational principle \cite{ripka, thouless}. In applications in light nuclei, the HF and J projection are known to account for 2-body correlation and give results close to shell model diagonalization \cite{macfarlane, khadkikar}. For heavy deformed nuclei, the model is known to give quantitative results for the band structures and deformations \cite{ghorui, naik, praharaj, patra, ghorui1, praharaj1, praharaj2, joshi, yokoyama, praharaj3, ghorui2}.

\paragraph{\bf{Hartree-Fock equations with axial symmetry:}}
The Hamiltonian consists of single-particle energy term and 2-body residual interaction:
\begin{eqnarray}
H=H_{S.P}+V
\label{eq:hf}
\end{eqnarray}
For axially symmetric HF field, an orbit is
\begin{eqnarray}
|\alpha m\big>=\sum_j C_{jm}^{\alpha} |jm\big>
\label{eq:hf}
\end{eqnarray}
The HF equations are
\begin{eqnarray}
(\epsilon_j-E_{\alpha})C_{jm}^{\alpha}+\sum_{j_1j_2j_4m_2}V(j_1mj_2m_2;jmj_4m_2)\nonumber\\ \rho_{j_4m_2j_2m_2} C_{j_1m}^{\alpha}=0
\label{eq:hf}
\end{eqnarray}
and
\begin{eqnarray}
\rho_{j_4m_2j_2m_2}=\sum_{\alpha(occupied)} C_{j_4m_2}^{\alpha}C_{j_2m_2}^{\alpha}
\label{eq:hf}
\end{eqnarray}
The HF Hamiltonian and self energy are:
\begin{eqnarray}
h=\epsilon+\Gamma
\label{eq:hf}
\end{eqnarray}
\begin{eqnarray}
\Gamma=V\rho
\label{eq:hf}
\end{eqnarray}
Equations (3) and (4) are solved by iteration for the amplitudes $C_{jm}^{\alpha}$ and energies of HF orbits $E_{\alpha}$. During the iteration procedure the residual interaction V is included in each step in building the HF Hamiltonian. The prolate HF solution for $^{182}$W is the lowest in energy and its orbits are shown in {\bf{Fig. 1}}. Some of the results of the HF solution (energies, quadrupole moments etc.) are given in {\bf{Table I}}. From {\bf{Table I}} one sees that the residual interactions among the nucleus (24 active protons and 26 active neutrons) are substantial. We emphasize that the quadrupole moments etc. follow dynamically from the HF solutions and are not externally imposed:
\begin{eqnarray}
\big<Q_{20}\big>_{HF}=Tr(Q_{20}\rho)
\label{eq:hf}
\end{eqnarray}
Thus, the converged HF solution provides a self-consistent deformed basis where the effect of residual interaction is included \cite{macfarlane, khadkikar}. Suitable intrinsic states are obtained as products of proton and neutron Slater determinants built from the HF orbits.

In the intrinsic state frame (the body frame) we have the deformed intrinsic state. This is not a state of unique angular momentum, rather it is a superposition of various J states. For rotational symmetry we need good angular momentum states. As pointed out by Peierls and Yoccoz (PY) \cite{peierls} the intrinsic frame is not conductive to states of good angular momentum. The PY procedure \cite{peierls} consists of rotating the intrinsic state (through Euler angles) and integration over the angles representing the orientation of the intrinsic frame. This is angular momentum projection.

\begin{table*}
\centering
\caption{Prolate HF solution of $^{182}W$ where substantial interaction energies are seen. Energies are in MeV. Mass quadrupole moment $Q_{20}$ and hexadecapole moment $Q_{40}$ are shown in power units of oscillator length parameter b. $b^2 \approx (0.9 A^{1/3}+0.7)$ fm$^2$ \cite{blomquist}.}
%\resizebox{0.7\textwidth}{!}{
%\renewcommand{\arraystretch}{2}
\scalebox{1.}{
\begin{tabular}{ccccccccccccc}
\hline\hline
 E$_{HF}$ &$\left< V_{pp}\right>$ & $\left< V_{pn}\right>$ & $\left< V_{nn}\right>$ &&  \multicolumn{2}{c}{Q$_{20}$ in $b^2$} & & \multicolumn{2}{c}{Q$_{40}$ in $b^4$}  \\ 
\cline{6-7}\cline{9-10}
        &&& &&       Proton & Neutron   &&     Proton & Neutron \\
   \hline
 -577.674&-33.222&-119.902&-43.475  &&  15.368&26.339 &&  -36.361&-52.920 \\
         
\hline\hline
\end{tabular}}
%}
\label{hforb1}
\end{table*}

\paragraph{\bf{Angular Momentum Projection formalism:}}
It is to be emphasized that an intrinsic state $|\phi_K\big>$, is a superposition of various angular momentum states:
\begin{eqnarray}
|\phi_K\big>=\sum_JC_{JK}|\psi_K^J\big>
\end{eqnarray}
The Peierls-Yoccoz procedure of angular momentum projection to a state of good angular momentum \cite{peierls,ripka} consists of applying a rotation operator $R(\Omega)$ ($\Omega$ stands for the Euler angles $\alpha$,$\beta$,$\gamma$) to  $|\phi_K\big>$ and integrating over the Euler angles with the D function:
\begin{eqnarray}
|\psi_K^J\big>=\frac{2J+1}{8\pi^2}\int d\Omega D_{MK}^{J*}(\Omega)R(\Omega)|\phi_K\big>
\end{eqnarray}
where, $R(\Omega)$ is the rotation operator $e^{-i\alpha J_z}e^{-i\beta J_y}e^{-i\gamma J_z}$.

Energies and electromagnetic transition operators are evaluated by calculating their matrix elements, which involves integration over Euler angles. There is axial symmetry of the Hartree-Fock field and the intrinsic states have good K quantum numbers. Thus, in the evaluation of the matrix elements between intrinsic states  $|\phi_{K_1}\big>$ and  $|\phi_{K_2}\big>$ the integration for two of the Euler angles $\alpha$ and $\gamma$ are done analytically and only the integration for Euler angle $\beta$ remains to be done numerically. We use 64 point Gauss-Legendre formula for the numerical integration of kernels over the angle $\beta$. Extended precision is used in the evaluation of the kernels. The matrix elements of the Hamiltonian and an electromagnetic operator of multipolarity L are:

\begin{widetext}
\begin{subequations}
\begin{align}
H_{K_1K_2}^J=\frac{(J+1/2)}{(N_{K_1K_1}^{J}N_{K_2K_2}^{J})^{1/2}}\int_0^{\pi}d\beta sin(\beta) d_{K_1K_2}^J(\beta)
 \big<\phi_{K_1}|He^{-i\beta J_y}|\phi_{K_2}\big> 
\end{align}

\begin{align}
\big<\psi_{K_1}^{J_1}||T^L||\psi_{K_2}^{J_2}\big>=\frac{(J_2+1/2)(2J_1+1)^{1/2}}{(N_{K_1K_1}^{J_1}N_{K_2K_2}^{J_2})^{1/2}}
\sum_{\mu(\nu)}C_{\mu\;\nu\;K_1}^{J_2\;L\;J_1}\int_0^\beta d\beta sin(\beta)
 d_{\mu K_2}^{J_2}(\beta)\big<\phi_{K_1}|T_{\nu}^L e^{-i\beta J_y}|\phi_{K_2}\big>
\end{align}
%\end{equation}
%\begin{equation}
where,
\begin{align}
N_{K_1K_2}^J=\int_0^\pi d\beta sin(\beta) d_{K_1K_2}^J(\beta)\big<\phi_{K_1}|e^{-i\beta J_y}|\phi_{K_2} \big>
\end{align}
%\end{equation}
\end{subequations}
\end{widetext}
is the overlap integral.

We mention here about the possibility of K-selection rule violating transitions in this theoretical formalism. The Clebsch-Gordan coefficient in Eqn.(10b) for the reduced matrix element contains the J selection rule for electromagnetic transitions and there is no K selection rule to prohibit a J-allowed transition. (As example, for E2 transition from J=10$^+$ K=10$^+$ to J=8$^+$ K=0$^+$, the coefficient $C_{2\;-2\;0}^{10\;2\;8}$ in Eqn.(10b) is non-zero and thus the transition is allowed by J selection rule). Because of the presence of the rotation operator $R(\Omega)$ in eqn.(10), there is a summation over $\mu (\nu)$ in the Clebsch-Gordan coefficient in Eqn.(10b). Thus, the angular momentum projection theory has J selection rule for the reduced matrix elements, but there is no K-selection rule in this theory.

Spectroscopic Quadrupole moment and Magnetic moment are
%\begin{widetext}
\begin{eqnarray}
Q_S(J)=\frac{1}{\sqrt{(2J+1)}} \sqrt{\frac{16\pi}{5}} C_{J\;0\;J}^{J\;2\;J}\big (\sum_{i_{(p,n)}} \big <\psi_K^J||Q_2^i|| \psi_K^J \big > \big )\;\;\;\;\;\;\;
\end{eqnarray}
\begin{eqnarray}
\mu (J)=\frac{1}{\sqrt{(2J+1)}} C_{J\;0\;J}^{J\;1\;J}\big (\sum_{i_{(p,n)}} \big <\psi_K^J||g_l^il_i+g_s^is_i|| \psi_K^J \big > \big ).\;\;\;\;\;\;\;
\end{eqnarray}
%\end{widetext}
\begin{table*}
\centering
\caption{BE(2) and BM(1) values for transitions from $K=10^+\rightarrow K=0^+$.}
\resizebox{0.6\textwidth}{!}{
%\renewcommand{\tabcolsep}{.03cm}
%\scalebox{1.2}{
\begin{tabular}{ccccccccccccc}
\hline\hline
 Transitions &&& BE(2) in $e^2 fm^4$ && & Transitions &&& BM(1) in $\mu_N^2$ \\

   \hline
 $10^+_{iso} \rightarrow 8^+_{gr}$&&&$0.4106\times10^{-4}$ &&&$10^+_{iso} \rightarrow 10^+_{gr}$&&&$0.3\times10^{-7}$\\
 $10^+_{iso} \rightarrow 10^+_{gr}$&&&$0.1963\times10^{-3}$ &&&$11^+_{iso} \rightarrow 10^+_{gr}$&&&$0.3070\times10^{-6}$\\ 
  $11^+_{iso} \rightarrow 10^+_{gr}$&&&$0.2353\times10^{-2}$&&&$11^+_{iso} \rightarrow 12^+_{gr}$&&&$0.4614\times10^{-6}$\\
 $12^+_{iso} \rightarrow 10^+_{gr}$&&&$0.8295\times10^{-2}$ &&&$12^+_{iso} \rightarrow 12^+_{gr}$&&&$0.7755\times10^{-5}$\\
 &&& &&&$13^+_{iso} \rightarrow 12^+_{gr}$&&&$0.3115\times10^{-4}$\\      
\hline\hline
\end{tabular}
}
\label{hforb2}
\end{table*}

%\begin{widetext}

%\end{widetext}

\begin{figure}[ht]
\begin{center}
\includegraphics[width=0.8\columnwidth]{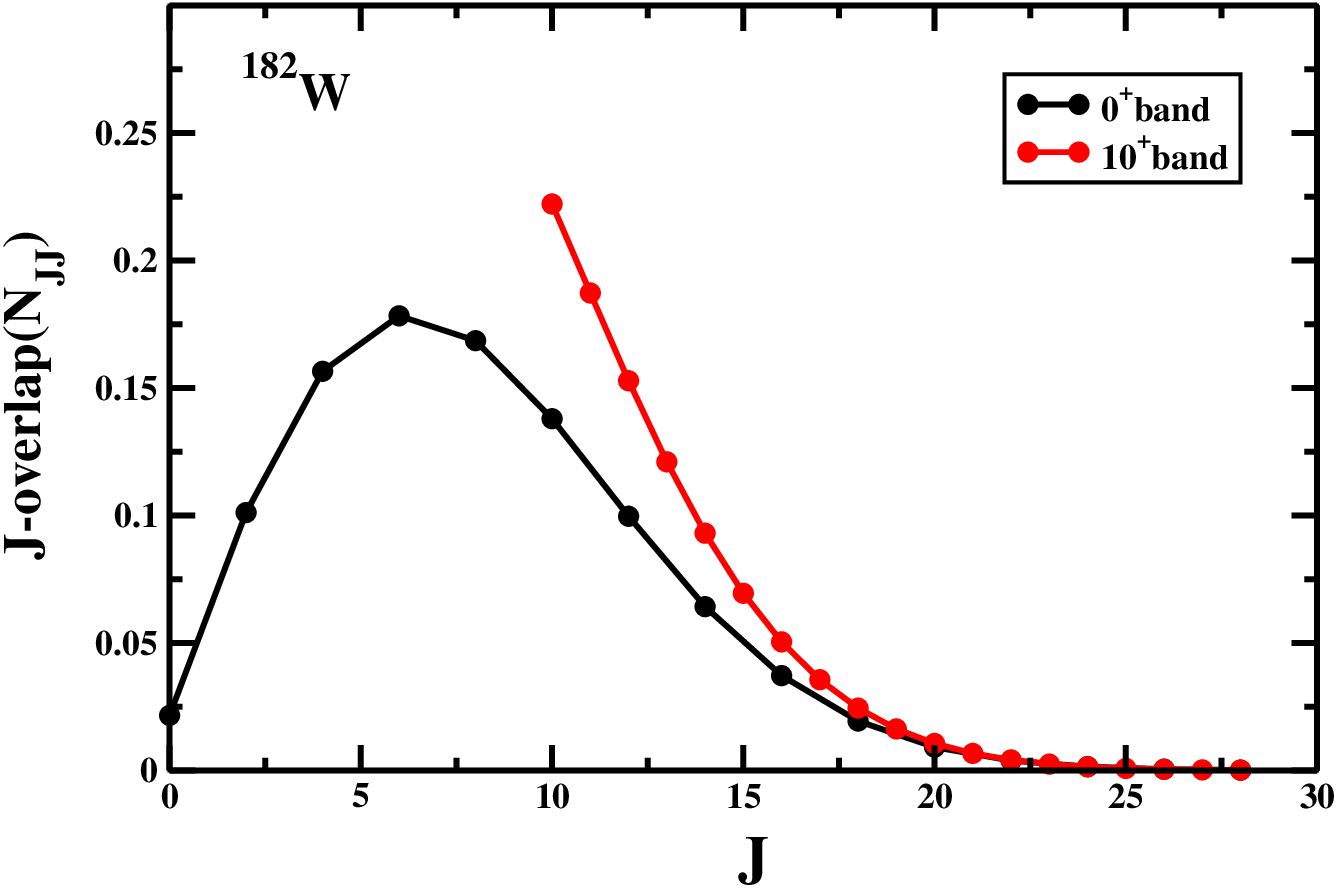}
\end{center}
\caption{Plot of J overlaps for K=0$^+$ and K=10$^+$ intrinsic states.}
\label{fig3}
\end{figure}

\paragraph{\bf{Results and Discussions:}}

The Hartree-Fock orbits of the prolate deformed solution of the Eqns.(3) and (4) are shown in {\bf{Fig. 1}} for $^{182}$W with 24 active protons and 26 active neutrons. The neutron orbit $\Omega$=9/2$^+$ is just below the Fermi level and the $\Omega$=11/2$^+$ orbit is above. One particle-one hole excitation across the neutron Fermi level gives K=10$^+$ configuration $\big (9/2^+n,11/2^+n\big)$. This configuration for K=10$^+$ band agrees with that in \cite{kondev}. The Hartree-Fock energy, pp, pn, and nn interaction energies and the deformation obtained in the HF solution are given in Table \ref{hforb1}. The interaction energies $\big <V_{pp}\big>$, $\big <V_{pn}\big>$, $\big <V_{nn}\big>$ are substantial in the deformed HF and the residual interactions are well taken into account \cite{macfarlane,khadkikar} in this theoretical procedure.

\begin{figure}[ht]
\begin{center}
\includegraphics[width=.8\columnwidth]{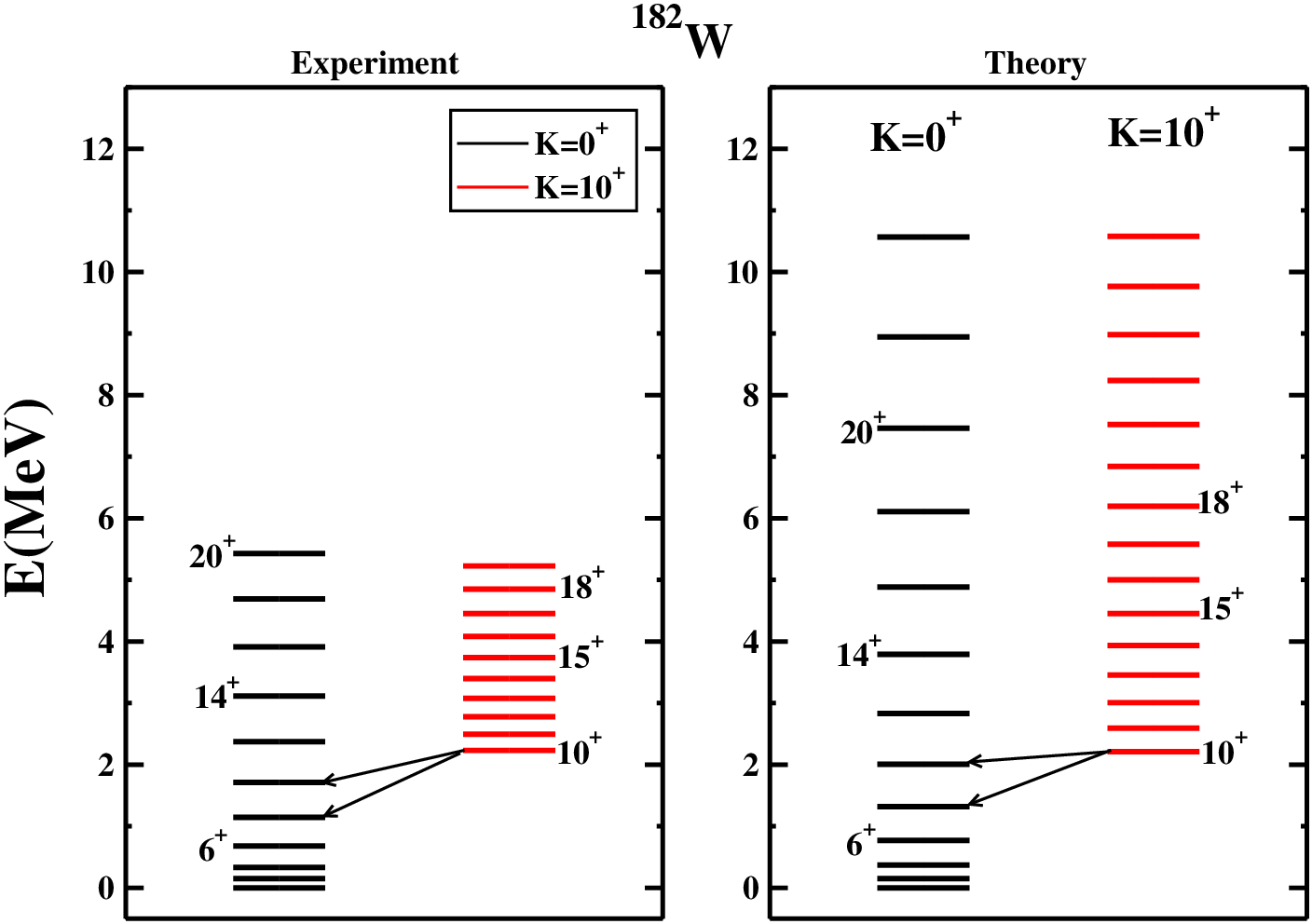}
\end{center}
\caption{Energy spectra of K=0$^+$ and K=10$^+$ bands of $^{182}$W obtained by J projection from respective intrinsic configurations. E2 and M1 transitions from the 10$^+$ band-head to the ground band are shown and compared with experimental data.}
\label{fig3}
\end{figure}

\begin{figure}[ht]
\begin{center}
\includegraphics[width=.9\columnwidth]{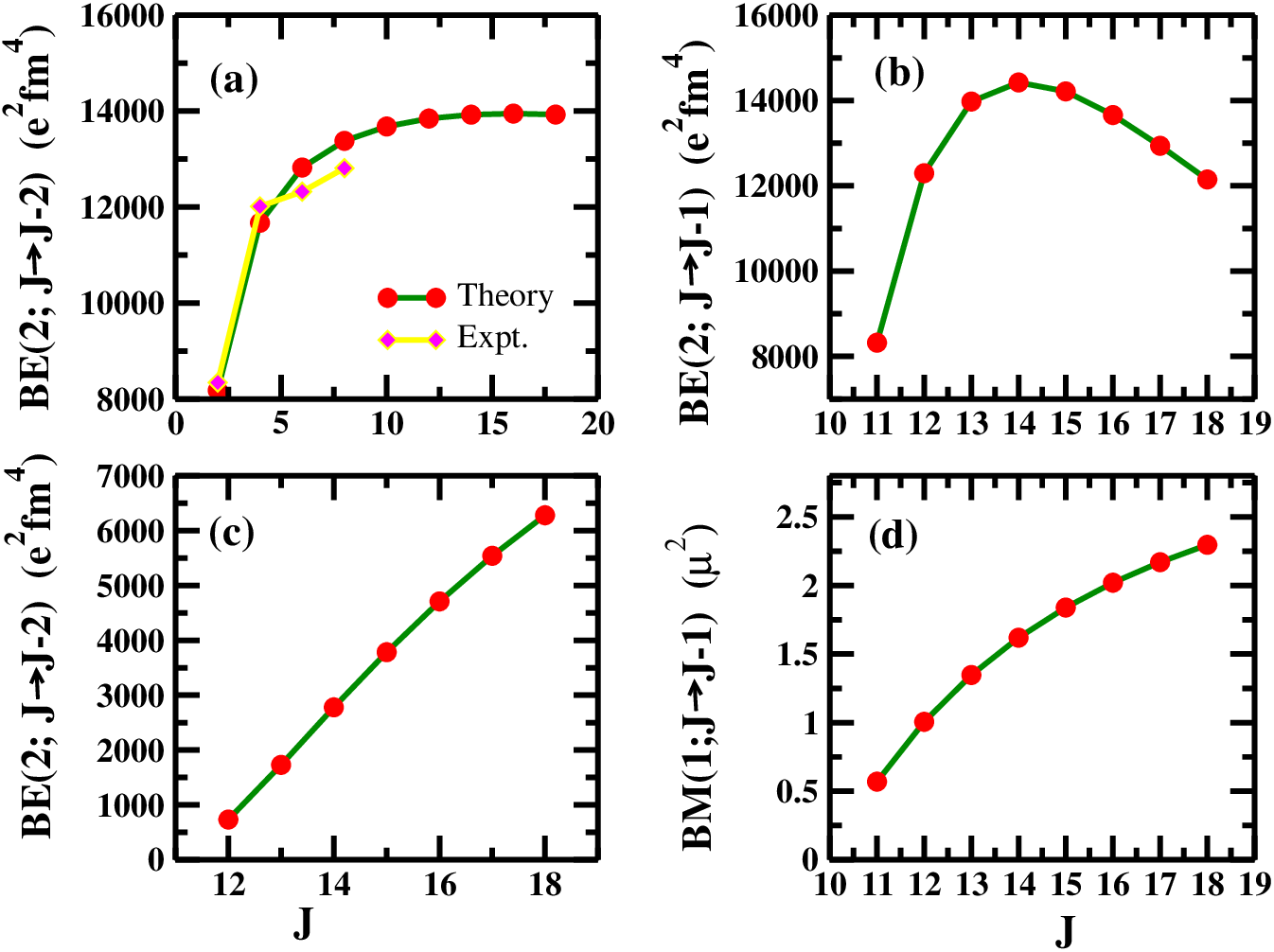}
\end{center}
\caption{(a) J to J-2 BE(2) values for K=0$^+$ band, (b) J to J-1 BE(2) values for K=10$^+$ band, (c) J to J-2 BE(2) values for K=10$^+$ band, (d) J to J-1 BM(1) values for K=10$^+$ band of $^{182}$W. Comparision is made with available experimental BE(2).}
\label{fig3}
\end{figure}
\begin{figure}[ht]
\begin{center}
\includegraphics[width=0.6\columnwidth]{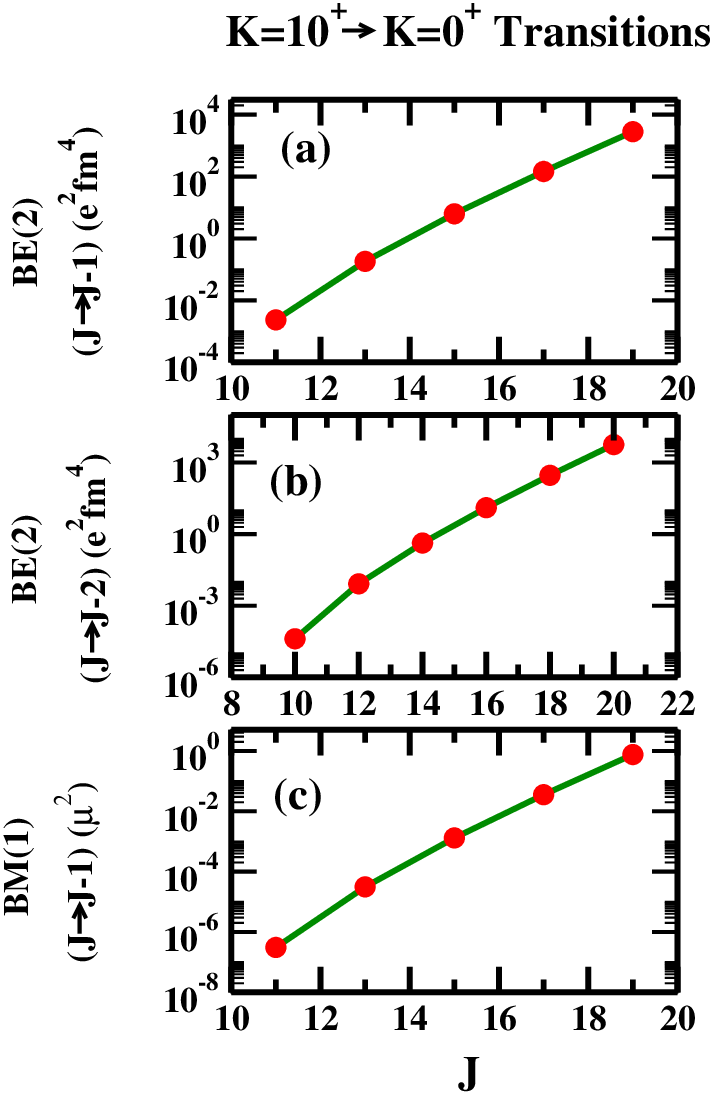}
\end{center}
\caption{Values of (a) BE(2)(J to J-1), (b) BE(2)(J to J-2), (c) BM(1)(J to J-1) for transitions from K=10$^+$ $\rightarrow$ K=0$^+$ of $^{182}$W.}
\label{fig3}
\end{figure}

\begin{figure}[ht]
\begin{center}
\includegraphics[width=0.6\columnwidth]{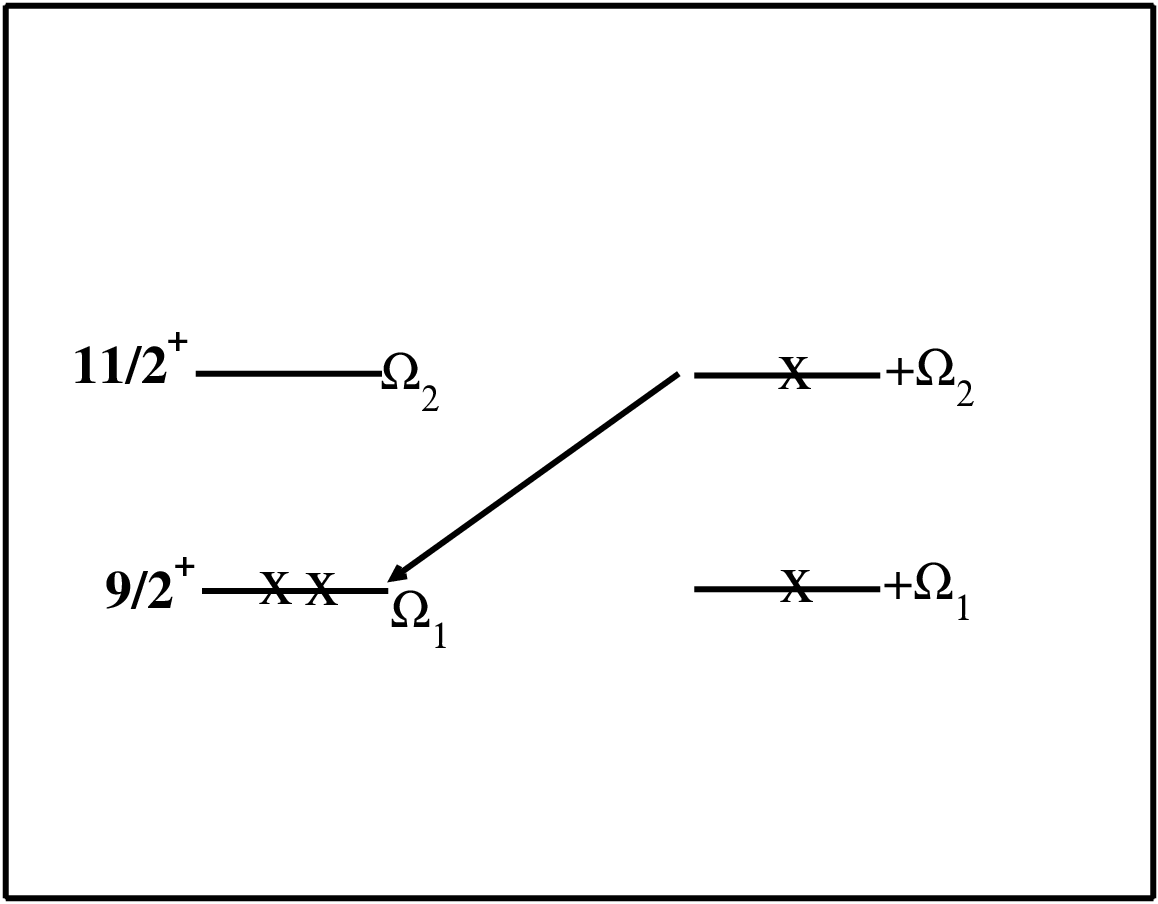}
\end{center}
\caption{Single nucleon transition in K-isomer decay ($\Omega_2$ to -$\Omega_1$). $\Omega_2$=11/2$^+$ and $\Omega_1$=9/2$^+$ for K=10$^+$ isomer of $^{182}$W.}
\label{fig3}
\end{figure}
We have done angular momentum projection from K=0$^+$ (ground band) and K=10$^+$ (isomer) intrinsic states and the results of the spectra are shown in {\bf{Fig. 3}}. The overlap integrals $N_{0^+0^+}$ and $N_{10^+10^+}$ Eqn.(10c) are shown in {\bf{Fig. 2}}, indicating well-spread wave-packets and these suggest the need for angular momentum projection to get states of good J. It is to be noted that the theoretical spectra shown in {\bf{Fig. 3}} have good K quantum numbers (K=0$^+$ and K=10$^+$).

%The overlap integrals $N_{KK}^J$ for the intrinsic states of the K=0$^+$ and K=10$^+$ bands are plotted in Fig. 2. The angular momentum spreads of the wave-packets are considerable, suggesting that angular momentum projection is essential in order to understand the finer details of spectra and transitions in deformed nuclei and their K-isomers unprojected wave packets will be far too crude.

%The Hamiltonian overlap between the ground band and the K-isomer band is
%$\big <10^+_{iso}|H|10^+_{gr}\big >=0.3419$ eV compared to $\Delta E\big (10^+_{iso}-10^+_{gr}\big)=0.519$ MeV. Thus there is practically no mixing between the K=0$^+$ ground band and the K=10$^+$ isomer band. Unlike in phenomenological models, K-mixing as a general phenomenon for explaining K-isomer decay is not there and the K quantum number is quite robust.

We have evaluated (Eqn. 10a) the energy overlap of the 10$^+$ states:
\begin{eqnarray}
\big <10^+_{isomer}|H|10^+{gr}\big>=0.342\;eV
\end{eqnarray}
whereas the experimental energy difference between the two 10$^+$ states is
\begin{eqnarray}
\Delta (10^+_{isomer}-10^+{gr})=0.519\;MeV.
\end{eqnarray}

Thus the interaction energy in Eqn.(13) is smaller than the energy difference by a factor of about 1.5$\times10^6$ and so there is practically no mixing between the two J=10$^+$ states of the K=0$^+$ and K=10$^+$ bands. This rules out the possibility of K mixing and thus the K quantum numbers of the two bands of $^{182}$W are quite robust.

\begin{table}
\centering
\caption{Spectroscopic Quadrupole Moments and Magnetic moments of K=0$^+$ and K=10$^+$ isomer bands from angular momentum projection.}
\resizebox{0.4\textwidth}{!}{
\begin{tabular}{ccccccccccccc}
\hline\hline
 Q$_S^{gr}$(2$^+$)&&$\mu_{gr}$ (2$^+$)&&Q$_S^{iso}$(10$^+$)&&$\mu_{iso}$(10$^+$)\\ 
(in e fm$^2$) &&(in $\mu$$_{N}$)&&(in e fm$^2$)&&(in $\mu_N$)\\
 \hline
-409.78&&1.494&&475.861&&-3.344 \\
% Q$_S^{gr}$(2$^+$) $\rightarrow$ &-431.81 e fm$^2$ \\
% $\mu_{gr}$ (2$^+$) $\rightarrow$ &1.558 $\mu_N$ \\
% Q$_S^{gr}$(10$^+$) $\rightarrow$ &501.337 e fm$^2$ \\ 
% $\mu_{iso}$(10$^+$) $\rightarrow$ &0.294 $\mu_N$  \\
\hline\hline
\end{tabular}
}
\label{hforb3}
\end{table}

The spectroscopic quadrupole moments and magnetic moments of the ground and K=10$^+$ bands are listed in {\bf{Table 3}}. In evaluating the magnetic moments and the BM(1) we use g$_l^p$=1.0, g$_l^n$=0.0, g$_s^p$=5.5857, g$_s^n$=-3.8261 and a spin quenching factor of 0.6.

In {\bf{Fig. 3}} the energy spectra of the ground band and the K=10$^+$ bandhead compare well with experiment. The level spacings of the K=10$^+$ band are a bit exaggerated in our model. In {\bf{Fig. 4 (a)}}, we plot the BE(2; J$\rightarrow $J-2) values for the K=0$^+$ band. In our calculations we have taken effective charges as 0.4 and 1.4 e for neuton and proton. The Weisskopf single-particle estimate for BE(2) for $^{182}$W is$\;\approx 61.3$ e$^2$fm$^4$ and we have BE(2; 2$^+$$\rightarrow$0$^+$) as 8181.4 e$^2$ fm$^4$ in our calculation, an enhancement of about 140 over single-particle value. The experimental BE(2; 2$^+$$\rightarrow$0$^+$) is 8342.93 e$^2$fm$^4$ \cite{raman}. Similarly, the experimental BE(2; 4$^+$$\rightarrow$2$^+$) =12014.8 e$^2$fm$^4$, BE(2; 6$^+$$\rightarrow$4$^+$) =12321.3 e$^2$fm$^4$ and BE(2; 8$^+$$\rightarrow$6$^+$) =12811.7 e$^2$fm$^4$ where as our theoretical BE(2) values for the same states are 11672, 12825 and 13379 e$^2$fm$^4$ respectively as shown in {\bf{Fig. 4(a)}}.

The BE(2) values of the K=10$^+$ band are plotted in {\bf{Fig. 4 (b)}} (J$\rightarrow$J-1) and in {\bf{Fig. 4 (c)}} (J$\rightarrow$J-2). They are collective, but the (J$\rightarrow$J-2) BE(2)'s are less collective of the two. The BM(1; J$\rightarrow$J-1) for the K=10$^+$ band are shown in {\bf{Fig. 4 (d)}}. The BE(2) and BM(1) values for transitions from the K=10$^+$ band to the K=0$^+$ ground band are given in {\bf{Table 2}} and {\bf{Fig. 5}}. These values for K=10$^+$$\rightarrow$K=0$^+$ transitions are finite and about $10^{-8}$ of the values of collective transitions within a band (Fig. 4) and are thus severely retarded. (For example, the BE(2) value for 10$^+_{gr}$ $\rightarrow$8$^+$$_{gr}$=13680 e$^2$ fm$^4$ and for 10$^+_{iso}$ $\rightarrow$8$^+$$_{gr}$=0.4106$\times$10$^{-4}$ e$^2$ fm$^4$). There is no K-selection rule in our theoretical model prohibiting  E2 and M1 transitions from K=10$^+$ to K=0$^+$ band- there is only J selection rule (Eqn. 10b). {\bf{Thus after careful calculation we find that the K-selection rule violating E2 and M1 transitions from the K=10$^+$ to K=0$^+$ band are finite, but retarded. This is in agreement with experimental trends \cite{kondev, regan, raman}.}} Our theory describes both the collective transitions within a band and the severely retarded (but non-vanishing) ones from K-isomers to lower K band.

\begin{figure}[ht]
\begin{center}
\includegraphics[width=.8\columnwidth]{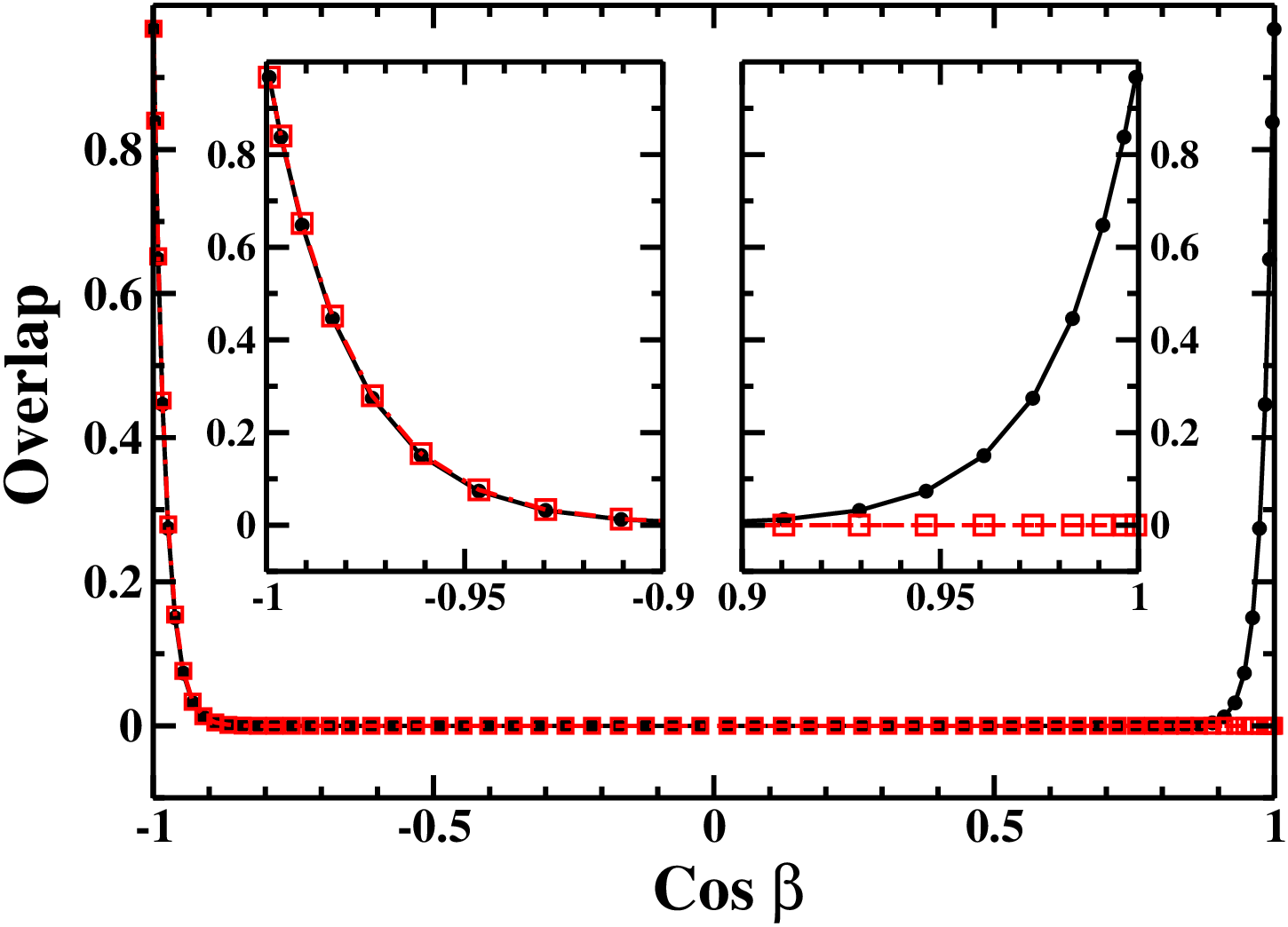}
\end{center}
\caption{Overlap kernels $\big<\phi_K|e^{-i\beta J_{y}}|\phi_K\big>$ of $^{182}$W for K=0$^+$ and K=10$^+$ intrinsic states. (Black dots) for K=0$^+$ and (red squares for K=10$^+$). }
\label{fig3}
\end{figure}

The overlap kernels $\big<\phi_K|e^{-i\beta J_{y}}|\phi_K\big>$ for the K=0$^+$ and the K=10$^+$ bands are plotted in {\bf{Fig. 7}}. K=0$^+$ configuration has time-reversal symmetry and its overlap kernel has very sharp peaks near $\beta$=0$^\circ$ and 180$^\circ$ falling off sharply within Cos$\beta$ interval of 0.05. The K=10$^+$ configuration has sharp peak near $\beta$=0$^\circ$ and falls off very fast with $\beta$ value. {\bf{Fig. 8}} shows the overlap kernel between K=0$^+$ and K=10$^+$ bands. The overlap kernel $\big<\phi_{0^+}|e^{-i\beta J_{y}}|\phi_{10^+}\big>$ is very small ($\sim$10$^{-8}$)and vanishes for $\beta$=0$^\circ$ and 180$^\circ$ and thus does not match with the kernels of either of the bands. The severe retardation of BE(2) and BM(1) values for transitions from K=10$^+$ to the ground band (Table 3) and the small overlap kernel between the two configurations ({\bf{Fig. 8)}} are reflections of the fact that there is a large change of single neutron configuration (11/2$^+$$\rightarrow$ -9/2$^+$) in the transition from K=10$^+$ to the ground band (depicted in {\bf{Fig. 6)}}. 

In the microscopic picture, a collective transition is caused by the coherent contributions of many nucleons. This is reflected in the fact that the overlap kernel is peaked near 0$^\circ$/180$^\circ$ in Euler angle $\beta$. In the case of K-isomer decay to lower K band only a single-nucleon contributes (and that by changing its $\Omega$ quantum number by a large amount, leading to further retardation). The process conserves angular momentum (rotational symmetry) for the combined system of nucleons and the $\gamma$-ray photon, given by the J-selection rule.

\begin{figure}[ht]
\begin{center}
\includegraphics[width=.8\columnwidth]{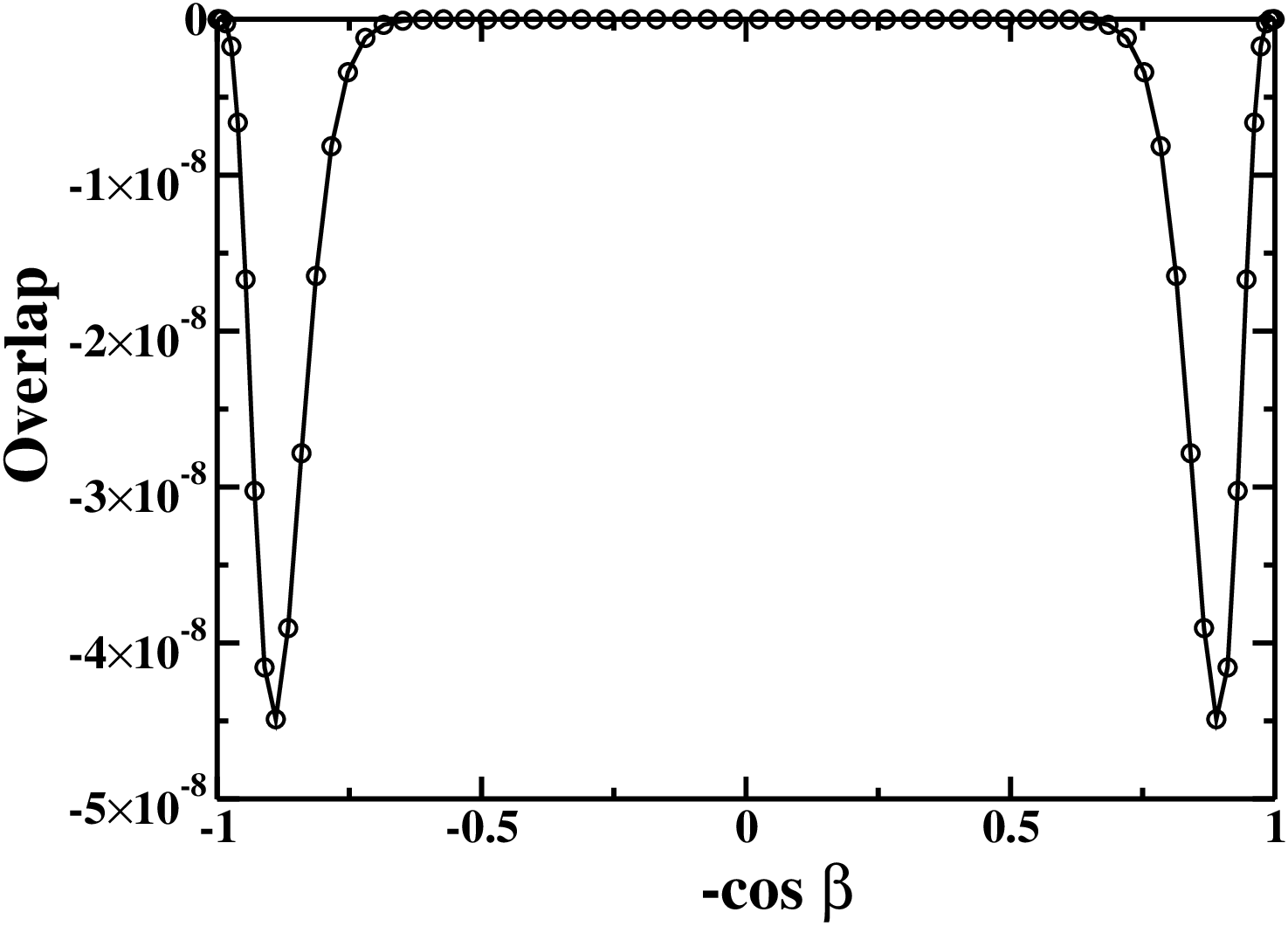}
\end{center}
\caption{K=0$^+$ and K=10$^+$ overlap kernel $\big<\phi_{0^+}|e^{-i\beta J_{y}}|\phi_{10^+}\big>$.}
\label{fig3}
\end{figure}

\paragraph{\bf{Interaction between the two bands and mixings:}}

The Hamiltonian overlap (interaction energy) between the K=10$^+$ and the  K=0$^+$ bands are plotted in {\bf{Fig. 9}} for various J values. It is seen that the interaction between the two bands are quite small at low J, (J=10$^+$ for example) and becomes quite higher as J increases. Thus, the mixing between the two bands is insignificant for K=10$^+$ bandhead and states just above. But at higher J the mixing can be more significant. Such large increase in interaction energy between the ground band and the isomeric band at higher J values may be a general feature of K-isomers for other nuclei too.

\begin{figure}[ht]
\begin{center}
\includegraphics[width=0.8\columnwidth]{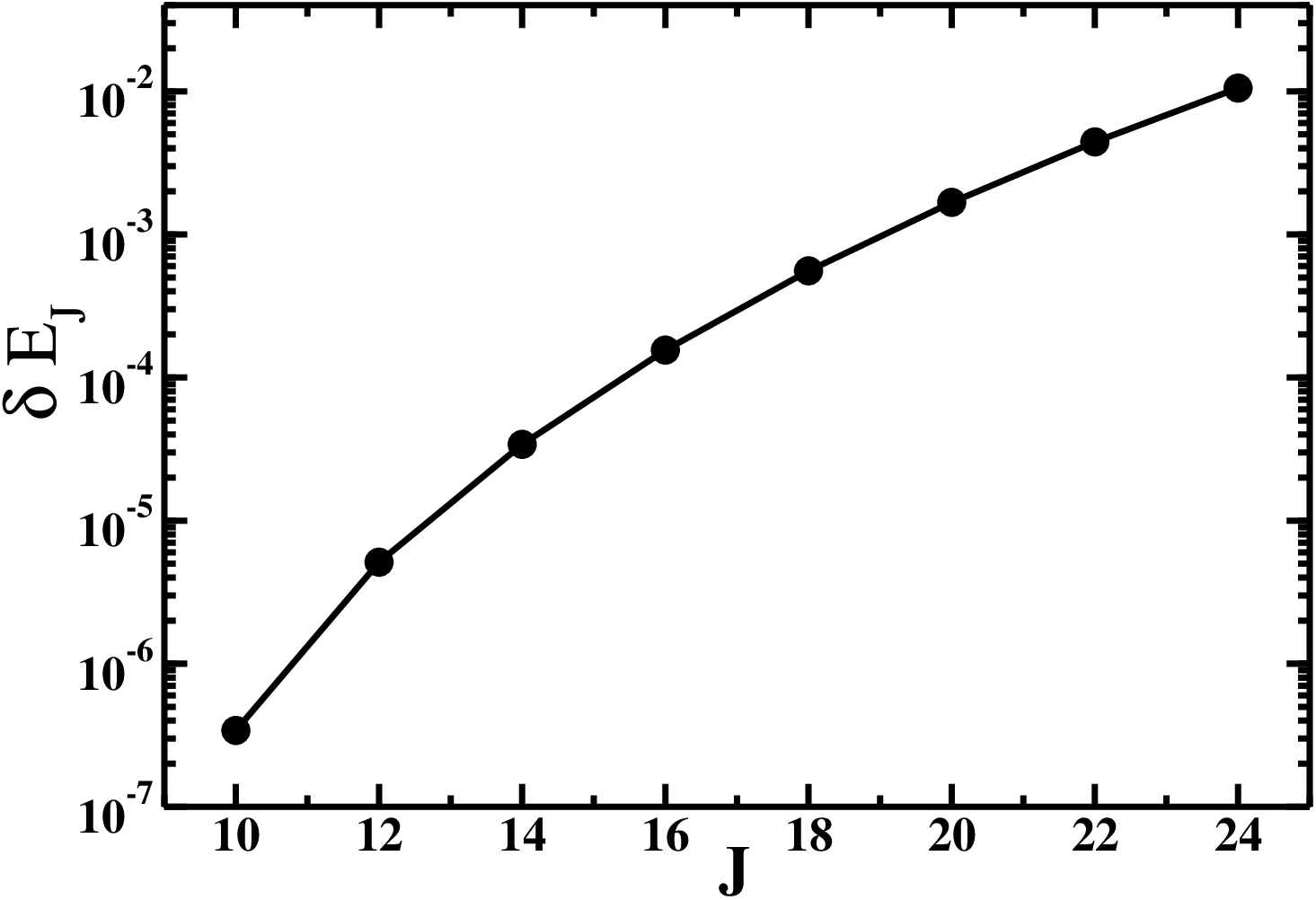}
\end{center}
\caption{Plot of interaction energy between the ground band and the K-isomer.}
\label{fig3}
\end{figure}

\paragraph{\bf{Summary and Conclusions:}}\label{conc}

We have shown in this work that one does not need K-mixing for transitions from K-isomer to lower K bands. Two K bands of good but widely different K values can be connected by low order multipoles, provided the J and parity selection rules are satisfied. It is essential to implement rotational symmetry by angular momentum projection from intrinsic states.

The spectra and transition matrix elements, including the K-selection rule violating E2 and M1 transitions from the K=10$^+$ bandhead of $^{182}$W to the ground band, have been studied by deformed Hartree-Fock and angular momentum projection. The spectra and the collectivity of transitions within bands and the retardations in K=10$^+$ to K=0$^+$ transitions are well reproduced. There is no K mixing. {\bf {The transitions allowed by J selection rule (J=10$^+_{iso}$$\rightarrow$8$^+_{gr}$ by E2 radiation, for example) are finite, but retarded, in agreement with experimental trends. The K quantum numbers of the respective bands are robust and K mixing is negligible in our microscopic calculation.}}

Thus, a theoretical understanding of the decay of a K-isomer to lower K bands is possible if rotational symmetry is implemented by angular momentum projection from the deformed intrinsic states. All transitions allowed by J and parity selection rules are in general possible. The retardation of such transitions from the K isomer is explained naturally by the poor overlaps between the K-isomer and the lower K configurations.

The constraint on the transition matrix elements imposed by the rotational model is violated in the microscopic model at the single-nucleon level (neutron $\Omega$=11/2$^+$$\rightarrow$ -9/2$^+$ for K=10$^+$ decay in $^{182}$W), in agreement with experimental trends. With this theoretical insight one can study the decay of K isomers in many other cases and this will help in the quantitative spectroscopy involving K-isomer for its many  possible applications.

\section*{Acknowledgments}
We thank Prof. R. K. Bhowmik, Prof. Partha Chowdhury and Prof. S. B. Khadkikar for discussions and suggestions. CRP was partly supported by SERB grant SB/S2/HEP-06/2013.

%\bibliography{mybibfile}
\end{document}